\documentclass[journal]{IEEEtran}
\usepackage{amsmath}
\usepackage{subfigure}
\usepackage{graphicx}
\usepackage{listings}
\usepackage{float}
\usepackage{stfloats}
\usepackage{upgreek}
\usepackage{blindtext}
\usepackage{enumitem}
\usepackage{booktabs}
\usepackage{multirow}
\usepackage{color}
\usepackage{graphicx}
\usepackage{cite}
\usepackage{picinpar}
\usepackage{flushend}
\usepackage{colortbl}
\usepackage{soul}
\usepackage{pifont}
\usepackage{alltt}
\usepackage{enumerate}
\usepackage{siunitx}
\usepackage{epstopdf}
\usepackage{pbox}
\usepackage[justification=centering]{caption}
\usepackage{threeparttable}
\usepackage[ruled]{algorithm2e}
\usepackage{url}
\usepackage{framed}

\usepackage{url}

\ifCLASSINFOpdf
\else
\fi
\begin{document}
%
\title{An ensemble learning approach for software semantic clone detection}
%
%
%

\author{Min~Fu,~\IEEEmembership{Member,~IEEE,}
        Gang~Luo,~\IEEEmembership{Member,~IEEE,}
        Xi~Zheng,~\IEEEmembership{Member,~IEEE,}
        Tianyi~Zhang,~\IEEEmembership{Member,~IEEE,}
        Dongjin~Yu,~\IEEEmembership{Member,~IEEE,}
        and~Miryung~Kim,~\IEEEmembership{Senior Member,~IEEE}
 \thanks{Min Fu is with the Alibaba Group, Hangzhou, China (e-mail: hanhao.fm@alibaba-inc.com).}

\thanks{Xi Zheng and Gang Luo are with the Department of Computing, Macquarie University, Sydeny, Australia.(e-mail: james.zheng@mq.edu.au (Corresponding Author); gangluo96@gmail.com).}
\thanks{Tianyi Zhang is with the Department of Computer Science, Harvard University, Cambridge, USA (e-mail: tianyi@seas.harvard.edu).}
\thanks{Dongjin Yu is with the Key Laboratory of Complex Systems Modeling and Simulation, Ministry of Education, Hangzhou Dianzi University, Hangzhou, China (e-mail: yudj@hdu.edu.cn).}
\thanks{Miryung Kim is with the Department of Computer Science, University of California, Los Angeles, USA (e-mail: miryung@cs.ucla.edu).}
}
\maketitle

\begin{abstract}
Code clone is a serious problem in software and has the potential to software defects, maintenance overhead, and licensing violations. Therefore, clone detection is important for reducing maintenance effort and improving code quality during software evolution. A variety of clone detection techniques have been proposed to identify similar code in software. However, few of them can efficiently detect {\itshape semantic clones}---functionally similar code without any syntactic resemblance. Recently, several deep learning based clone detectors are proposed to detect semantic clones. However, these approaches have high cost in data labelling and model training. In this paper, we propose a novel approach that leverages word embedding and ensemble learning techniques to detect semantic clones. Our evaluation on a commonly used clone benchmark, BigCloneBench, shows that our approach significantly improves the precision and recall of semantic clone detection, in comparison to a token-based clone detector, SourcererCC, and another deep learning based clone detector, CDLH.
\end{abstract}

\begin{IEEEkeywords}
software clones, clone detection, random forest, word embedding.
\end{IEEEkeywords}

%
\IEEEpeerreviewmaketitle

\section{Introduction}
\IEEEPARstart{S}{oftware} developers often implement functions based on existing ones or simply even reuse code via copy\&paste, which generates many syntactically or functionally similar code---{\itshape code clones}. Though such code reuse practices can significantly improve development efficiency, code clones often lead to software defects, maintenance overhead, and even licensing violations. As software systems are increasing in size and complexity, clone detection becomes increasingly important and crucial during software evolution.

Many approaches and tools have been proposed to detect code fragments with similar syntax or functionalities. These approaches can be mainly grouped into three categories: text-based, token-based, and tree-based. NICAD~\cite{roy2008nicad} is a text-based code clone detection approach. It first uses flexible pretty-printing and code normalization to preprocess code, and then dynamically clusters potential clones with simple text-line comparison. CCFinder~\cite{kamiya2002ccfinder} and SourcererCC~\cite{sajnani2016sourcerercc} are two popular token-based approaches. They treat code as bags of tokens and calculate the similarity between different token sequences. DECKARD~\cite{jiang2007deckard} is a tree-based clone detector that identifies similar code in terms of Abstract Syntax Trees (ASTs). Although these tools are pragmatic to use and achieves good scalability to a large code corpus, they only rely on syntactic similarity to identify similar code and therefore cannot efficiently detect {\itshape semantic clones}---functionally similar code without much syntactic resemblance.

Recently, several clone detection approaches have been proposed to leverage deep learning models to detect semantic software clones, i.e., Type 4 clones according to a well-known clone taxonomy~\cite{davey1995development, roy2009comparison}. CCLEARNER is the first token-based clone detection approach that leverages deep learning~\cite{li2017cclearner}. It extracts tokens from source code and represents them as feature vectors. Then it applies Deep Neural Network (DNN) to identify code with similar features. CDLH is another tree-based deep learning approach~\cite{wei2017supervised}. It uses AST-based Long Short-Term Memory (LSTM) to exploit the lexical and syntactical information from code and achieves good performance for Type-3 and Type-4 clone detection. Though CCLEARNER and CDLH achieve better precision and recall than previous unsupervised clone detectors, there are still rooms for improvement in terms of training and prediction efficiency and clone detection accuracy. Also, they do not adequately scale to big code, since deep learning models cost significant training time.

In this paper, we propose a scalable and effective approach for semantic clone detection. Our approach leverages word embedding to exploit the lexical and syntactical information from program source code and transforms a code fragment into a feature vector~\cite{zhang2017service2vec}. Instead of using sophisticated deep learning models, our approach leverages ensemble learning, which is proven to be more efficient and scalable. To evaluate our approach, we use F1-score as an evaluation metric~\cite{kim2008classifying,rahman2012recalling,nam2013transfer}.
We perform experiments on a widely used clone benchmark, \textit{BigCloneBench}~\cite{svajlenko2014towards}, which is commonly used to evaluate existing clone detection techniques~\cite{wei2017supervised}.  We compare our approach with the state-of-the-art baseline, i.e., \textit{CDLH}, the approach proposed by Wei et al~\cite{wei2017supervised}. The experiment results show that our approach achieves better performance than the state-of-the-art approaches in terms of F1-score. In particular, in the most difficult Type-4 clone detection, our approach achieves a F1-score of over 96\%, precision of over 99\% and recall of over 92\%, which are significantly higher than the baseline. We achieve this improvement by maximally maintaining the lexical and syntactical information in the original code fragment pairs through our word embedding based feature engineering, and using a classification learning model to leverage all features in the input vectors.

The contributions of our study are:
\begin{itemize}
\item[1)] We propose a novel approach, which leverages word embedding, text/document embedding, and ensemble learning techniques in tandem to achieve better performance.
\item[2)] We propose a feature engineering approach for analyzing source code and code pairs, which minimizes the information lost. Compared to the state of the art deep learning clone detector \textit{CDLH} on the popular public dataset \textit{BigCloneBench}, our approach achieves better performance for Type-4 clone detection.
\end{itemize}


In the paper, the background of our study, the overall framework and elaborates of our approach, the experiments and results are respectively introduced in Section~\ref{sec:background}, Section~\ref{sec:algorithm} and Section~\ref{sec:exp}. And the related work is discussed in Section~\ref{sec:related}. The last section is the Conclusion.

\section{Preliminaries}\label{sec:background}

We first present the four types of software clones in the section. Next, we introduce the background of word embedding and ensemble learning used in our approach.

\subsection{Software Clone Types}\label{sub:sct}


Generally, software clone can be divided into 4 types based on degrees of similarity:

\begin{itemize}
  \item [1)] {\bf Type-1.} Two code fragments are exactly the same except for comments and layouts.
  \item [2)] {\bf Type-2.} In addition to Type-1 clone differences, the two code fragments are different in identifier names and literal values.
  \item [3)] {\bf Type-3.} In addition to Type-1 and Type-2 clone differences, one of the two code fragments was revised, added or deleted relative to the other one. In other words, they are syntactically similar in the statement level.
  \item [4)] {\bf Type-4.} Two code fragments implement the same functionality, while they are not similar syntactically.
\end{itemize}

\subsection{Word Embedding Techniques}\label{sub:wet}


Since 1954, many different DSMs were proposed. In DSMs, to make the vector representation of words appearing in similar contexts similar, each word is embedded in a real-valued d-dimensional vector. Now, with the deep learning techniques developing, some effective neural network models are proposed~\cite{bengio2006neural,collobert2008unified,mikolov2013efficient,mikolov2013distributed}. These models can generate a low-dimensional word vector representations by using DNN to learn from the context of the corpus. The method automatically generates useful low-dimensional representations is called {\itshape word embedding}. For various information retrieval tasks, it has been proven that the perform of embedding is better than traditional count-based models~\cite{ye2016word,chen2016mining}.

\begin{figure}
    \centering
    \includegraphics[width=7cm]{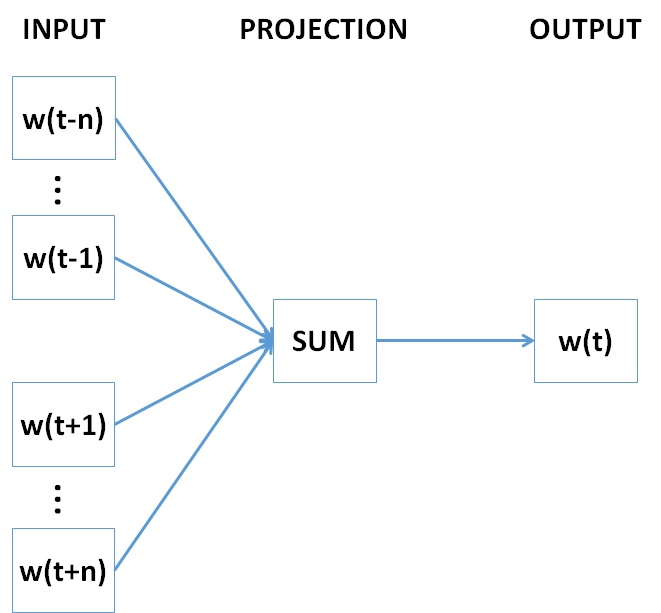}
    \caption{~The architectures of CBOW models.}
    \label{fig:cbow}
    \vspace{0cm}
\end{figure}

\begin{figure}
    \centering
    \includegraphics[width=7cm]{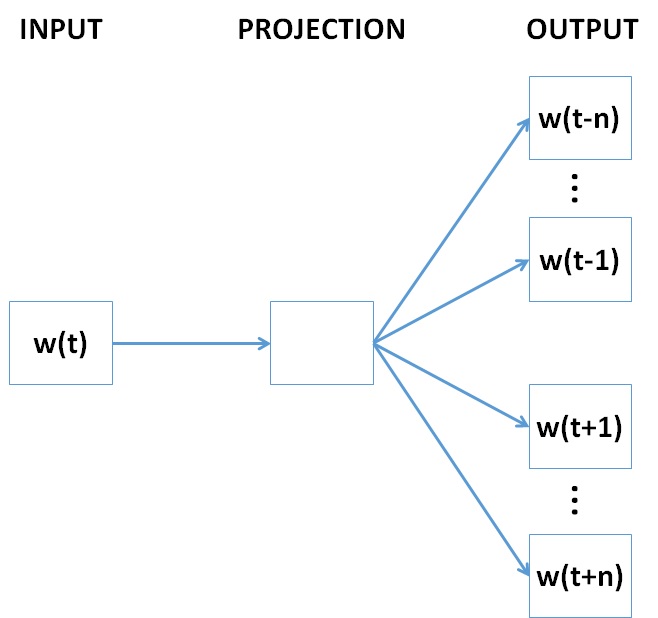}
    \caption{~The architectures of skip-gram models.}
    \label{fig:skip}
\end{figure}


There are two widely used embedding models. One is CBOW model and the other one is skip-gram model~\cite{mikolov2013efficient,mikolov2013distributed}. The figures~\ref{fig:cbow} and~\ref{fig:skip} respectively show the architecture of CBOW model and the skip-gram model. In the figure, the $w(t)$ denotes the current word and the $w(t \pm i)$ (i=1,2,...,n) is the surrounding words of the current word. Specifically, the CBOW model gives a prediction of the current word based on its context. For the skip-gram model, the difference with the CBOW model is that it gives a prediction of the surrounding words based on its current word. In prior work on solving other software engineering tasks the skip-gram model had been proven working well.(e.g., log analysis and code annotation)\cite{ye2016word,chen2016mining,fu2020ics}. So it be considered using in our study.

\subsection{Ensemble Learning}\label{sub:el}


Learning involved multiple learners is increasingly popular ~\cite{deng2020dynamic,zhang2020achieving,zhang2020democratically}, and among them ensemble learning is simple and effective ~\cite{polikar2012ensemble}. In general, there are many different characteristics in different classifiers in terms of sensitivity to different training data and intrinsic principles. One of the key elements of ensemble learning is base learners. For base learners, supporting vector machine classification techniques and decision tree classification techniques can be chosen~\cite{han2006data}. The other key elements of ensemble learning is ensemble methods. Similarly there are many effective ensemble methods, such as stacking, bagging and boosting~\cite{Charu2015data}.

In the paper, all techniques of classification can be used to base learners. In our problem, as we think in normal code each kind of token is definitely followed by several kinds of tokens, code fragment generally then can be modeled as a tree-like structure with each node represents a token or a feature variable. Decision Tree can find the feature variables rapidly. And the feature variables are the best for differentiating different classes. Besides, for different classes, many other classifiers cannot generate explicit rules, but the feature variables can generate it~\cite{han2006data}. Therefore, the decision tree method is a proper choice. In theory, as long as the decision tree is large enough, it can cover all cases. In the preliminary experiment, we also verify that the decision tree can achieve the best performance among the other classifiers (e.g., Naive Bayes, Support Vector Machine, Linear Discriminant Analysis, and Nearest Neighbour Classifier).


As mentioned above, to handle all cases of token subsequence, we may build a very huge decision tree, which leads to too much time and space cost and is not practical. Therefore, we leverage ensemble learning, expecting to build a number of medium-size decision trees in replace of a huge one. Specifically, we use the bagging to create a random forest. Bagging, also known as bootstrapped aggregation, may minimize prediction variance\cite{Charu2015data}. In bagging, as the data of the original training data set is uniformly sampled and replaced, different sampled data sets may generate different models. Finally, in different models, the class with the majority vote will be regarded as the final predictive model.



%
\section{Our Proposed Approach}\label{sec:algorithm}

We first introduce the individual steps of our approach in the section. Next, we explain the details of the framework.

\subsection{Overall Framework}\label{sec:of}



\begin{figure}
    \centering
    \includegraphics[width=8cm]{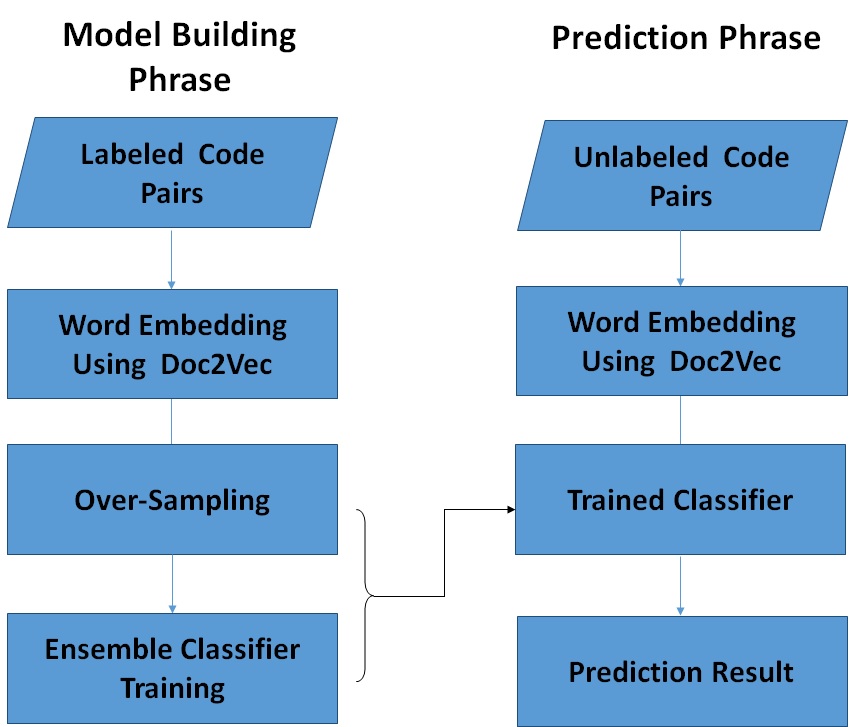}
    \caption{~The Overall Framework of Our Approach.}\vspace{0cm}
    \label{fig:frame}
\end{figure}

The framework of our approach is presented in Figure~\ref{fig:frame}. Code clone detection, especially type 4 clone detection, can be performed better by analyzing existing code pair sets and extracting those features which maintain the original code information than most recent deep learning approaches (e.g., using AST-based LSTM~\cite{wei2017supervised}) which extract a feature matrix from the original code to fit into the input requirement of deep learning models. We argue our approach can perform significantly better both in accuracy and testing time than deep learning models due to much less loss of information in feature engineering and using less expensive classification models. We also argue our approach has better accuracy than the previous approach in clone detection due to leveraging intuitive feature engineering (i.e., Doc2Vec~\cite{mikolov2013distributed}) and a powerful ensemble classifier (i.e., Random Forest).


Generally, there are two stages to the framework: the phase of model building and the phase of prediction. Our aim in the process of model building is based on the decision trees to construct an ensemble classifier. This ensemble classifier will determine in the predictive process whether an unknown code pair would be a clone or a non-clone.

Firstly, a lot of features are extracted from a training set of code pairs in our framework (i.e., code pairs with known labels). The features of a code pair consist of two 100-dimensional vectors, and each vector represents a corresponding code fragment. In this paper, we use a word embedding technique, i.e., Doc2Vec~\cite{mikolov2013distributed} (an unsupervised algorithm to generate vectors for sentence/paragraphs/documents) to extract features from code\footnote{Details in Section~\ref{sec:we}.}.


Then, we build the prediction model by using bagging and decision tree. Before training the model, we firstly perform over-sampling\footnote{Details in Section~\ref{sec:os}.} to dealing with class imbalance~\cite{he2009learning}. This is because the negative class data in the dataset is much less than the positive class data. To make the negative cost of the model training equivalent to the positive cost of the model training, balancing the data with over-sampling is required, and this can lead to a more accurate classification training model. Then, the balanced dataset is used to train an ensemble classifier - Random Forest\footnote{Details in Section~\ref{sec:el}.}.


Code pair with unknown labe is predicted to be clone or non-clone by the trained ensemble classifier during the prediction phase. For each code pair, the same set of features is extracted by using Doc2Vec during the model construction phase. Then input these features into the trained Random Forest classifier. In the end, the classifier produces a prediction result, which is one of the following labels: a clone or a non-clone.

\subsection{Word Embedding: Word2Vec}\label{sec:we}


The word embedding technique is crucial in this paper, we leverage the skip-gram model ~\cite{mikolov2013efficient,mikolov2013distributed} to transform each code fragment into a vector. Specifically, if given a token $t$ in code (e.g., a word appearing in the code), we can use the $C_{t}$ to represent the set of the surrounding context tokens (e.g., the neighboring words of the given token). Given the objective function $J$ of a skip-gram model is the sum of log probabilities of the surrounding context tokens under the condition of a given token:

 \[
    J=\sum_{i=1}^{n} \sum_{t_{j} \in C_{t_{i}}} \log p(t_{j}|t_{i})
 \]


In the formula above, $n$ represents the full length of the token sequence. And the $p(t_{j}|t_{i})$ denotes the conditional probability defined as following:

 \[
    p(t_{j} \in C_{t_{i}}|t_{i})=\frac{exp(v_{t_{j}}^{T}v_{t_{i}})}{\sum_{t \in T}exp(v_{t}^{T}v_{t_{i}})}
 \]


In the formula above, $v_{t}$ denotes the vector form of the word $t$. $T$ denotes the vocabulary of all tokens. By training an entire corpus of code fragments, a $d$-dimensional vector can represent all tokens in the vocabulary of the corpus where $d $is a variable parameter and is set to an integer (e.g., $90$).

As a general practice, we can set the value of $d$ to be equal to the size of the maximum vector among all code vectors. In our case, the maximum vector size happens to be $100$. We experiment with different sizes and find $100$ is the most appropriate for our training set. Choosing a vector size that is too big or small could lead to overfitting or underfitting.


Each token is transformed into a fixed-length vector by using the skip-gram model. Theoretically, each row of the matrix represents a token, and the entire matrix can denote a code fragment. However, it is difficult to feed them directly into the predictive model as input since different code fragment has different numbers of tokens. Therefore, by averaging all the token vectors the code fragment contains, we transform the code fragment matrix into a vector. The average is calculated based on the numeric value of each dimension in the vector. Specifically, given a code fragment matrix that has $n$ rows in total, we denote the i-th row of the matrix as $r_{i}$ and the transformed code fragment vector $v_{d}$ is generated as follows:

 \[
    v_{d}=\frac{\sum_{i}r_{i}}{n}
 \]

With the above formula, each code fragment can be represented as a feature vector, which can represent the characteristic of a code fragment.

\subsection{Imbalanced Learning: Over-Sampling}\label{sec:os}

In order to perform the model training, we need to prepare the training data set. Since there are original $250,000$ items of negative class data (non-clone pairs) in the data set, we need to generate more negative class data so that the negative data can be spread across all possible negative cases of code clones.

For the clone pairs inside the BigCloneBench, the true clone and false clone pairs are all manually labelled beforehand. We convert these clone pairs into vectors by using the doc2vec library thus generating our vector features.

With feature vectors, we can then build prediction models. However, the initial training dataset is very unbalanced (the amounts of labelled clone pairs and non-clone pairs are rather unequal), which could heavily influence the performance of prediction models. Therefore, it is necessary to balance the training dataset by useing imbalanced learning strategies before building a model.


In previous studies, the two common imbalanced strategies are under-sampling and over-sampling~\cite{han2006data,he2009learning}. In our study, we prefer to consider more different code fragments, since more different code fragments have more different cases of token sequences. Therefore, over-sampling is a better choice. Over-sampling obtains a promising result on deal with the class imbalance task~\cite{han2006data,he2009learning}. For expanding its scale, this approach duplicates data which is belonging to the minority class. Generally, over-sampling sets the target ratio $p$, which is the ratio of data instances which is belonging to the minority class to the whole training datasets. Before the ratio of the minority class to the whole training data is greater than or equal to $p$, repeating the following two steps:


\vspace{0.1cm}\noindent{\itshape Step 1: The Selection of instance.}
There are variable strategies to select instance such as random selection and cluster-based selection. Choosing one method to select an instance belonging to the minority class. We choose random selection because each code pair in the minority class is independent of one another.



\noindent{\itshape Step 2: The Addition of instance.}
Add the instance selected in step 1 into the training dataset to expand its scale.


In our work, the amount of instances belonging to the minority class occupies almost equal to ten percent of the number of instances belonging to the majority class, so we simply duplicate all instances belonging to the minority class times to balance the amount of two classes.

\subsection{Ensemble Learning: Bagging}\label{sec:el}

With the balanced training data, we build a prediction model. At the same time, to construct a Random Forest model, we combine decision trees and bagging~\cite{Charu2015data}. The random forest integrates randomness into the model-building process of each decision tree to make the bagging work best.

%
%
\section{Experiments and Results}\label{sec:exp}

The effectiveness of our approach is evaluated in this section. The experimental environment as follows: The CPU is Intel(R) Core(TM) T6570 and its frequency is 2.00GHz; The RAM is 8GB, and the operation system is Windows 7. In Section~\ref{sec:setup} we present our dataset. In Section~\ref{sec:metric} we present our experiment setup. Next, in Section~\ref{sec:rqs}, we present our research questions and our experiment results. 

\subsection{Dataset: BigCloneBench}\label{sec:setup}

We evaluate our approach on the popular dataset \textit{BigCloneBench}~\cite{svajlenko2014towards}, which is also used by many software clone detection projects~\cite{sajnani2016sourcerercc,wei2017supervised}. \textit{BigCloneBench} consists of clones of common functionalities in IJaDataset. In \textit{BigCloneBench}, the clones are classified into $4$ types based on the syntactical similarity of clones, which is measured using a line-based metric.

Specifically, two code fragments can be categorized as Type-1 if they become textually identical after Type-1 normalization, which includes removing comments and strict pretty-printing. Similarly, two code fragments can be typified as Type-2 if they become textually identical after Type-2 normalization, which expands Type-1 normalization to include systematic renaming of identifiers, and replacing of literals with default values (e.g., numeric to 0, strings to default). If two code fragments are not identical after these normalizations yet if their functionalities are the same, their syntactic similarity are measured and they are categorized as Type-3 or Type-4 according to the syntactical similarity value. In particular, if the syntactical similarity value is in [0.7, 1), the clones are Strongly Type-3; if the syntactical similarity value is in [0.5, 0.7), the clones are Moderately Type-3; if the syntactical similarity value is in [0, 0.5), the clones are Weakly Type-3 or Type-4.

In addition, \textit{BigCloneBench} also contains many false clone pairs.
BigCloneBench dataset covers several business functionalities and we use 10 representative ones out of them (e.g., Web Download, Copy a File, Bubble Sort). The BigCloneBench dataset also contains 8 million true clone pairs in total, and over 6.1 million of them are related to the 10 representative business functionalities. We use them as the true clone pairs. We also use over 250 thousand false clone pairs in the BigCloneBench dataset. Table~\ref{tab:dataset} summarizes the statistics of \textit{BigCloneBench} in detail.

\begin{table*}[hb!]
    \fontsize{9pt}{\baselineskip}\selectfont
    \centering
    \caption{~Statistics of BigCloneBench.}
    \label{tab:dataset}
    \begin{tabular}{l l l l l l l}
        \hline\hline
        \textbf{Functionality} & \textbf{Type-1} & \textbf{Type-2} & \textbf{Strongly Type-3} & \textbf{Moderately Type-3} &\textbf{Weakly Type-3 or Type-4} & \textbf{False}\\
        \hline
        Web Download & 1,554 & 9 & 1,439 & 2,715 & 410,611 & 38,838  \\
        Secure Hash (MD5)  & 632 & 587 & 3,294 & 24,923 & 871,717 & 4,564 \\
        Copy a File     & 13,805 & 3,116 & 5,947 & 24,199 & 4,725,438 & 204,108\\
        Decompress Zip  & 0 & 0 & 1 & 1 & 34 & 56\\
        FTP Authenticated Login & 9 & 0 & 94 & 191 & 49,161 & 4,202\\
        Bubble Sort & 43 & 4 & 239 & 1,752 & 13,538 & 5,432\\
        Init. SGV With Model & 3 & 7 & 5 & 2 & 259 & 78\\
        SGV Selection Event Handler & 0 & 0 & 0 & 0 & 55 & 1,272\\
        Create Java Project (Eclipse) & 0 & 0 & 8 & 0 & 245 & 0\\
        SQL Update and Rollback & 122 & 10 & 259 & 97 & 8,828 & 24\\
        Total & 16,168 & 3,733 & 11,286 & 53,880 & 6,079,886 & 258,574\\
        \hline\hline
    \end{tabular}
\end{table*}

We set the parameters of the random forest training model as specified below: tree number=100; tree depth=infinite; min sample for leaf node=2; mean ratio between the leaf node and the parent node=0; random data size for each tree=100000; random seed=0. These parameters are determined based on the full analysis of all training data in the BigCloneBench set as well as the engineered features originating from this set.

We use the ground truth provided by BCB to calculate the precision, instead of manually sampling and inspecting detected clone pairs. Even though BCB may miss some true clones in the ground truth, our calculated precision should only be lower than the true precision of our tool.

From Table~\ref{tab:dataset}, we can see that \textit{BigCloneBench} is very imbalanced. The number of true clone pairs are much more (over $20$ times) than that of false clone pairs, and most (98.6\%) of all are Weakly Type-3 or Type-4 clones. Therefore, an imbalanced learning strategy shall be adopted in the data preprocessing before the model training step.

In the paper, we denote T1 as Type-1 clones, T2 as Type-2 clones, ST3 as Strongly Type-3 clones, MT3 as Moderately Type-3 clones and T4 as Weakly Type-3 or Type-4 clones for simplicity.


\subsection{Experiment Setup}\label{sec:metric}

In software clone detection, we focus more on detected clones than non-clones, and we expect both high precision and high recall for detected clones. Therefore, we use Precision, Recall and F1-score to measure performance of our approach.

We used ten-fold cross-validation when we evaluated the performance of our approach~\cite{han2006data}. We concluded the average performance from 100 times running results of ten-fold cross-validation, so that we can make the experiment results more convincing.

\subsection{Research Questions}\label{sec:rqs}

The first baseline is an unsupervised learning approach \textit{SourcererCC} proposed by Sajnani et al~\cite{sajnani2016sourcerercc}. \textit{SourcererCC} is token-based and it leverages an optimized partial index and filtering heuristics to achieve large scale clone detection. SourcererCC is mainly designed for detecting Type1\&2 code clones; for Type 3\&4 code clones, it has limitations with detection because SourcererCC is based on tokenization of source code pairs and lacks the semantic information of each code snippets. The second baseline is a deep learning approach \textit{CDLH} proposed by Wei et al~\cite{wei2017supervised}. \textit{CDLH} uses AST-based LSTM to mine lexical and syntactical information from code and maps the information into Hamming Space using a hash function so that it achieves good performance especially for Type-4 clones. 

\vspace{0.1cm}\noindent{\textbf{RQ1  }} \textbf{How effective is our approach?}

\noindent{\bf Motivation.} The more an approach can detect correct clone pairs, the more effective the approach is. Therefore, we first evaluate the effectiveness of our approach against the two baselines.


\noindent{\bf Approach.} We use the precision, recall and F1-score, to make comparisons. In addition, we report the results respectively for different clone types, i.e., T1, T2, ST3, MT3 and T4.


\noindent{\bf Results.} Tables~\ref{tab:precision}, \ref{tab:recall} and \ref{tab:f1} present the Precision, Recall and F1-score values of our approach against the baselines.

\begin{table}[b]
    \centering
    \caption{~Precision of our approach and the two baselines.}
    \label{tab:precision}
    \begin{tabular}{l l l l}
        \hline\hline
        \textbf{Types} & \textbf{SourcererCC} & \textbf{CDLH} & \textbf{Our Approach} \\
        \hline
        T1   & 1.00 & 1.00 & 1.00 \\
        T2    & 1.00 & 1.00 & 1.00 \\
        ST3        & 0.38 & 0.90 & 1.00  \\
        MT3        & 0.15 & 0.86 & 1.00  \\
        T4        & 0.04 & 0.78 & 0.99  \\
        \hline\hline
    \end{tabular}
\end{table}

\begin{table}[b]
    \centering
    \caption{~Recall of our approach and the two baselines.}
    \label{tab:recall}
    \begin{tabular}{l l l l}
        \hline\hline
        \textbf{Types} & \textbf{SourcererCC} & \textbf{CDLH} & \textbf{Our Approach} \\
        \hline
        T1   & 1.00 & 1.00 & 1.00 \\
        T2    & 1.00 & 1.00 & 1.00 \\
        ST3        & 0.32 & 1.00 & 1.00  \\
        MT3        & 0.09 & 1.00 & 1.00  \\
        T4        & 0.01 & 0.81 & 0.928  \\
        \hline\hline
    \end{tabular}
\end{table}

\begin{table}
    \centering
    \caption{~F1-score of our approach and the two baselines.}
    \label{tab:f1}
    \begin{tabular}{l l l l}
        \hline\hline
        \textbf{Types} & \textbf{SourcererCC} & \textbf{CDLH} & \textbf{Our Approach} \\
        \hline
        T1   & 1.00 & 1.00 & 1.00 \\
        T2    & 1.00 & 1.00 & 1.00 \\
        ST3        & 0.36 & 0.94 & 1.00  \\
        MT3        & 0.13 & 0.88 & 1.00  \\
        T4        & 0.02 & 0.81 & 0.963  \\
        \hline
    \end{tabular}
\end{table}

From these tables, we can find that in terms of precision, recall and F1-score, our approach performs the best in all types, especially in type-4 clone detection. For SourcererCC, it has good performance in type-1 and type-2 clone detection, but it performs very poorly in type-3 and type-4 clone detection. The reason is that the tokenization-based transformation of the source code used by SourcerCC only deals with the text-level and structure-level similarity of the code, and it does not represent semantical information underlining the code. For CDLH, it also has good performance in type-1 and type-2 clone detection, while it largely improves the performance in type-3 and type-4 clone detection. In particular, it achieves a F1-score of 0.81 in type-4 clone detection. However, our approach can detect almost all clone types. A simple ensemble learning model can have surprisingly better accuracy in detecting type-4 clones than a complicated deep learning model (LSTM) by using more native and suitable features.

\begin{figure*}[ht!]
    \centering
    \includegraphics[width=16cm]{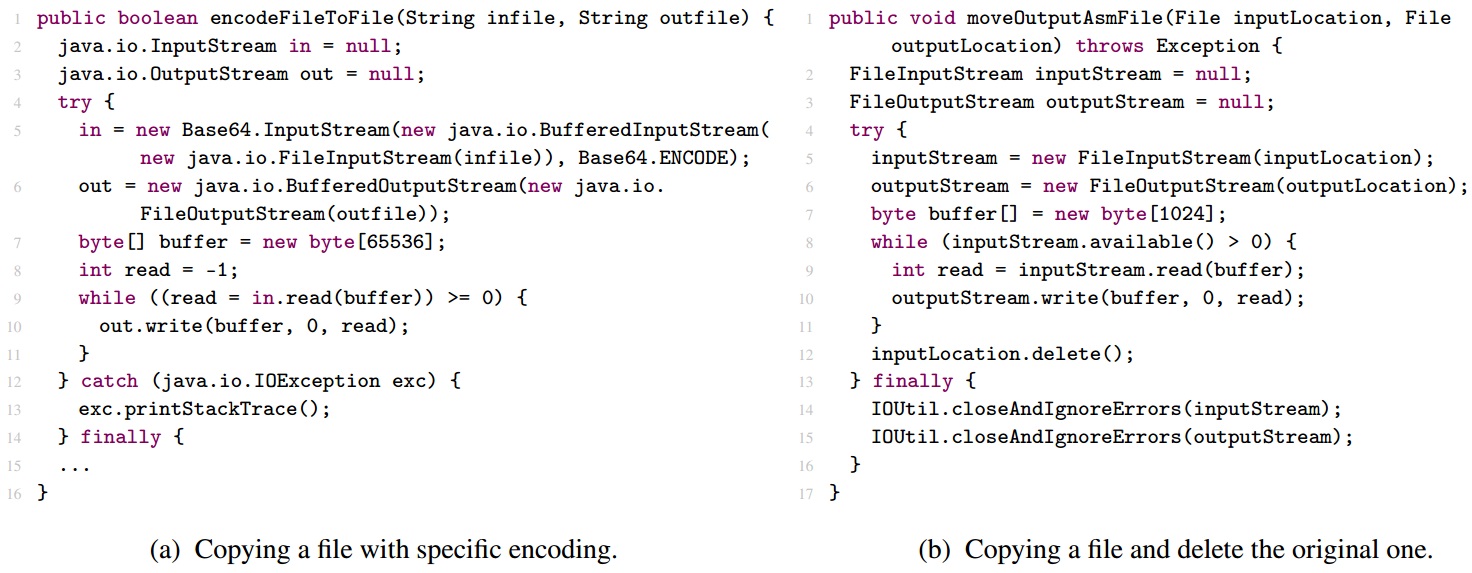}
    \caption{~A true positive pair of semantic clones.}
    \label{fig:tp}
    \vspace{0cm}
\end{figure*}

\begin{figure*}[ht!]
    \centering
    \includegraphics[width=16cm]{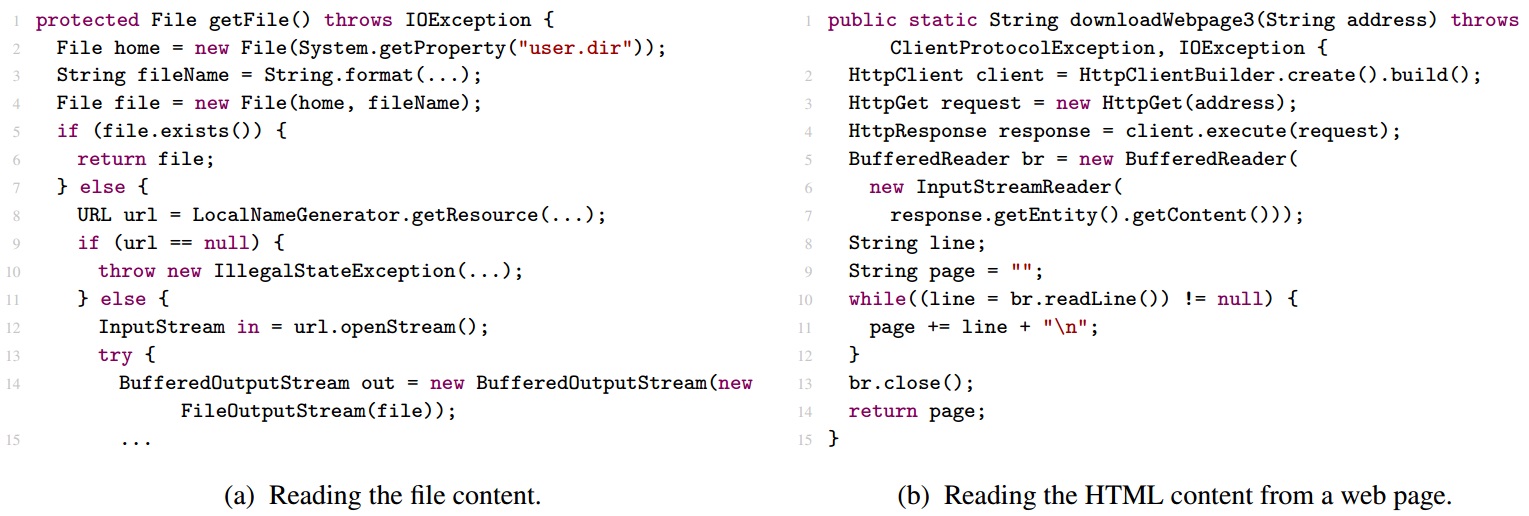}
    \caption{~A false positive pair of semantic clones.}
    \label{fig:fp}
    \vspace{0cm}
\end{figure*}

\begin{figure*}[ht!]
    \centering
    \includegraphics[width=16cm]{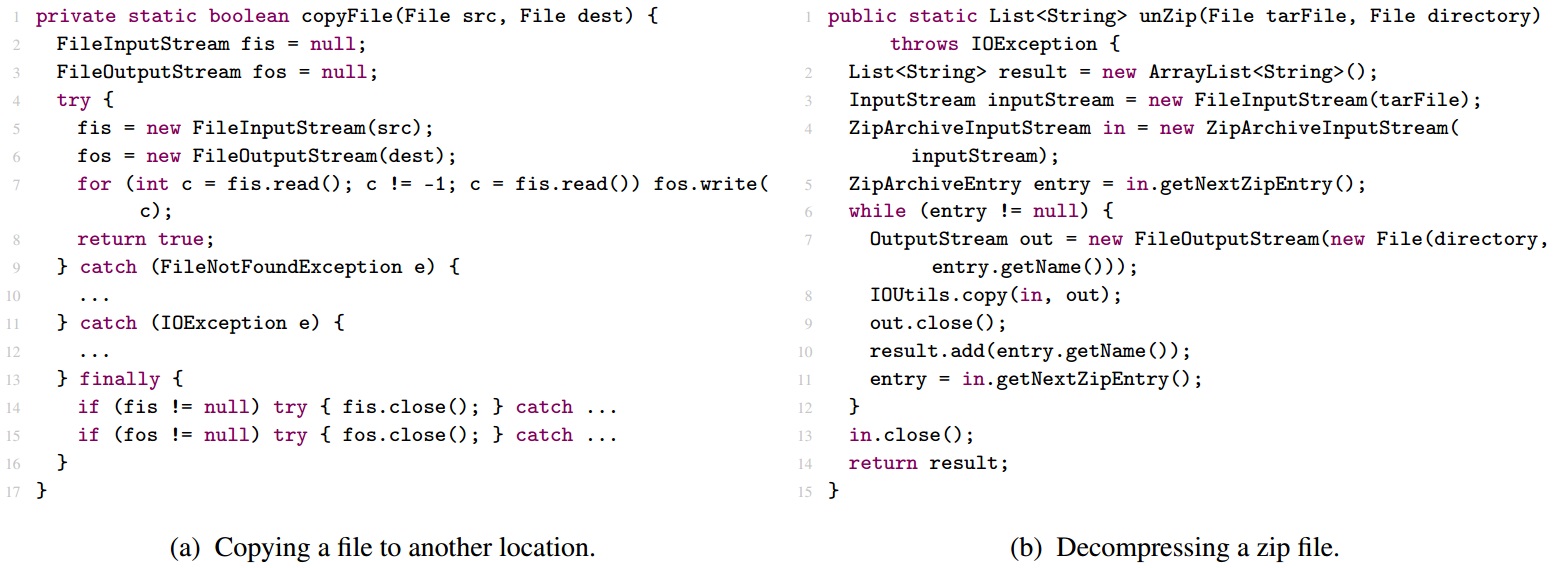}
    \caption{~A false negative pair of semantic clones.}
    \label{fig:fn}
    \vspace{0cm}
\end{figure*}

Figure~\ref{fig:tp} shows an example pair of semantic clones labelled in \textit{BigCloneBench}. This pair of semantic clones is detected by our approach but fails to be detected by SourcererCC and CDLH. In Figure~\ref{fig:tp}, both Java methods accomplish similar functionality by copying file content from file to another, although there is behaviour variation---one using a specific encoding character set, {\ttfamily Base64} while the other removing the original file after copying its content. SourcererCC fails to detect this pair since there is little overlap between their tokens. CDLH fails to detect it since its model only leverages syntactic information extracted from ASTs but does not incorporate any semantic information about the function's behaviour. Figure~\ref{fig:fp} shows two methods that are mistakenly detected as semantic clones by our approach. Though both methods have read operations, the left method retrieves the content from a given file and writes it to another file, while the right one reads HTML content from a given web address to a string. From the Doc2Vec point of view, the generated feature vectors are too similar to distinguish from each other. This indicates a future direction of incorporating code comments or annotations to feature vectors so that we can distinguish code with similar low-level operations yet different high-level behaviours. Figure~\ref{fig:fn} shows a pair of semantic clones that are missed by our approach. The reason is that due to the insufficient training set of semantic clones that use different data structures. Our random forest classifier has not been adequately trained to understand that the introduction of new data structures such as {\ttfamily List<String>} and {\ttfamily IOUtils} in the right method, which does not significantly impact its semantic behaviour. In our future work, we need to consider using a more advanced classification model to learn this kind of knowledge from the existing training set. Nevertheless, our approach is generally more effective than other baselines.


We computed a p-value by executeing Wilcoxon signed-rank statistical test and calculated the delta of the Cliff to better demonstrate the benefits of our approach. The Cliff's delta is in the range [-1, 1], where -1 or 1 respectively indicates all values in one group are less than or greater than the values in another group, and 0 indicates that the data in both groups are similar. Table 6 shows the correspondence between Cliff's delta scores and effectiveness levels. We can rigorously assess the extent of improvement of our approach over the two baselines by calculating the p-value and the Cliff's delta. In Table~\ref{tab:rank}, the range of ($\delta$) denotes the value of Cliff's Delta. And the value of effectiveness levels 1 to 4 respectively represent Negligible, Small, Medium and Large. 


\begin{table}
    \centering
    \caption{~Cliff's Delta values effectiveness correspondence table~\cite{cliff2014ordinal}.}
    \label{tab:rank}
    \begin{tabular}{l l}
        \hline\hline
        \textbf{Range of ($\delta$)} & \textbf{value of effectiveness levels} \\
        \hline
        $\delta$ $\in$ [-1, 0.147) & 1 \\
        $\delta$ $\in$ [0.146, 0.33) & 2 \\
        $\delta$ $\in$ [0.33, 0.474) & 3 \\
        $\delta$ $\in$ [0.474, 1] & 4 \\
        \hline
    \end{tabular}
\end{table}

\begin{framed}
\textit{Our approach is more effective than the baselines for software clone detection. In particular, our approach can achieve a F1-score of 96.3\% in Type-4 clone detection.}
\end{framed}

\vspace{0.1cm}\noindent{\textbf{RQ2  }} \textbf{How much time does it take for our approach to run?}


The training time and clom detection time of our approach are measured. The training time consists of the time spent in word embedding, over-sampling, and classifier training. The clone detection time is equal to the time from preprocessing the testing dataset to generating the predicted result. For software clone detection, there can be much committed code everyday and the number of pairing can be very big. Therefore, both the training and clone detection time are important. The training time should ensure that the model can be updated every day, and the clone detection time should ensure that all pairs of daily committed code can be detected as clones or non-clones.

%
%


\noindent{\bf Results.} We compare the training time of our approach against that of CDLH and SourcererCC. We takes about 6 hours to complete the training of the statistical model, while SourcererCC needs $1.7$ hours for data preprocessing and CDLH needs over one day for training. SourcererCC is the fastest since it is an unsupervised learning approach and does not really have a comparable training phase. CDLH is the lowest since it leverages RNN, which has high time complexity. Therefore, CDLH is not practical for daily software clone detection, and our approach is comparable to SourcererCC.

We also compare the clone detection time of our approach against that of CDLH and SourcererCC. It takes about $1.6$ hours for our approach to predict unknown clone pairs, while SourcererCC needs $0.4$ hours and CDLH still needs over one day. We can see that our approach is also comparable to SourcererCC in terms of clone detection time and can be practical. It also further backs up our finding that a simple ensemble learning model can be more practical for software clone detection than a complicated deep learning model (LSTM). The key insight is that training one classifier to detect all different types of clones is inadequate. Ensemble learning overcomes this limitation by integrating multiple classifiers for more robust prediction. Using Word2Vec to convert source code to vectors omits the syntax structures in source code but still memorizes important semantic information such as variable names and function calls, which is critical to detect Type 4 clones. The main reason why the LSTM approach does not work well for Type 4 clones is that it generates code embedding from ASTs, which focuses more on the structure of code. However, Type 4 clones often do not share similar code structures.

\begin{framed}
\textit{Our approach needs around $6$ hours to build a statistical model and $1.6$ hours to predict unknown clone pairs. Therefore, it can be used in practice.}
\end{framed}

\subsection{Threats to Validity}

We have evaluated our approach on over $6$ million true clone pairs and false clone pairs from \textit{BigCloneBench}. In the next work, we plan analyzing other datasets to reduce this threat further. We use precision, recall, F1-score to evaluate the effectiveness of various prediction techniques~\cite{kim2008classifying,rahman2012recalling,nam2013transfer,canfora2013multi,rahman2013sample,rahman2013and,peters2013better}.

\section{Related Work}\label{sec:related}
Software cloning is a major contributor to the steadily increasing complexity of software systems. 7\% to 23\% of source code in modern software systems are code clones~\cite{baker1995finding, Al-Ekram2005:byaccident, roy2008empirical}. Prior work has studied code cloning through both automated~\cite{baxter1998clone} and ethnographic~\cite{kim2004ethnographic} approaches. Kim et al.~find that developers often apply similar edits to code clones, but code clones could also easily diverge from one another, which may lead to program inconsistencies or bugs~\cite{kim2005empirical}. Chou et al.~also find that many operating system bugs are introduced by editing mistakes between clones~\cite{chou2001empirical}. The existence of code clones could also increase software maintenance effort, since developers need to update similar code distributed across multiple locations consistently. Such edits on code clones are hard to comprehend and inspect during peer code reviews\cite{dunsmore2000object, zhang2015interactive}. Therefore, detecting and managing code clones are often considered necessary to reduce software maintenance effort and improve software quality.


Generally, the code clones are categorized into four common types as described in Section~\ref{sub:sct} based on the degree of variants.


Over past decades, various clone detectors have been proposed to detect different types of clones. Text-based clone detectors are language independent by treating programs as plain texts~\cite{johnson1993identifying, manber1994finding, rieger2005effective, lee2005sdd, simian}. However, text-based approaches are known to be limited to detecting Type-1 and Type-2 clones only.


Token-based clone detection can detect Type-2 clones well by parameterizing identifiers and constant values~\cite{kamiya2002ccfinder, baker2007finding, li2006cp}. Token-based techniques utilize a lexical analyzer to tokenize source code and identify parameter tokens (i.e., identifiers and constant values), based on the grammar of the programming language. Parameter tokens are then replaced with abstract symbols. However, since syntax is not taken into account, clones found by token-based techniques may be invalid syntactic units.

Syntax-based clone detection is more robust than token-based clone detection by accounting for program syntax. Syntax-based techniques parse source code to syntax trees and then identify similar subtrees using tree matching algorithms~\cite{baxter1998clone, yang1991identifying, koschke2006clone, evans2009clone} or using metric-based clustering algorithms.\cite{jiang2007deckard, davey1995development} For example, Koschke et al.~\cite{koschke2006clone} serializes AST subtrees into node sequences and identify similar subsequences using suffix tree matching. Metric-based techniques extract syntactic metrics (e.g., the number of loops and branches) from syntax trees and then cluster metric vectors that have similar values. For example, Deckard encodes syntax trees to numerical vectors and clusters these vectors based on their Euclidean distance~\cite{jiang2007deckard}. Deckard optimizes the vector clustering algorithm using locality sensitive hashing (LSH). LSH can generate the same hash value for vectors within a given Euclidean distance.





To better detect Type-3 and Type-4 clones, some supervised learning methods have been proposed~\cite{li2017cclearner,wei2017supervised}. Li et al.~present CCLEARNER, which is the first solely token-based clone detection approach leveraging deep learning~\cite{li2017cclearner}. CCLEARNER extracts tokens from known code clones and non-clones as features and uses Deep Neural Network (DNN) to train the model. Wei et al.~propose an end-to-end deep feature learning framework called CDLH for functional clone detection~\cite{wei2017supervised}. CDLH uses AST-based LSTM to mine the lexical and syntactical information from code and maps the information into the Hamming Space using a hash function. With the technique, CDLH achieves good performance in Type-3 and Type-4 clone detection. Supervised learning clone detectors can often have better performance than traditional token or syntax based approaches, especially in Type-3 and Type-4 clone detection. However, since they have a training process, they cost much more time to detect clones. In addition, they need many known clones and non-clones, which may cost much time and effort to label manually.

\section{Conclusion and Future Work}\label{sec:conclusions}

In the paper, a nover approach to detect software clons is proposed. The approach leverages word embedding and ensemble learning techniques. To evaluate the approach, we use a commonly-used dataset \textit{BigCloneBench}, which contains over six million true clone pairs and false clone pairs. The experiment results show that our approach is better than the state-of-the-art approaches including CDLH using deep learning models and the improvement is statistically significant.


In the future work, we will focus on improving the performance of our approach by exploring different ensemble learning methods and different classifiers. We also plan to explore featuring engineering to substantiate our finding that a simple ensemble learning model can outperform complicated deep learning models by merely adopting better and more suitable features, which are proven efficient in other fields ~\cite{lu2019mfe,fu2018dalim}.


%




\ifCLASSOPTIONcaptionsoff
  \newpage
\fi



%
\bibliographystyle{IEEEtran}
\bibliography{references}

\end{document}